\documentclass[pra,twocolumn,superscriptaddress,a4paper]{revtex4}

\usepackage{graphicx}
\usepackage{dcolumn}
\usepackage{bm}
\usepackage{amsmath,amssymb}

\begin{document}

\newcommand{\be}{\begin{equation}}
\newcommand{\ee}{\end{equation}}
\newcommand{\der}[2]{{\frac{d #1}{d #2}}}
\newcommand{\beq}{\begin{eqnarray}}
\newcommand{\eeq}{\end{eqnarray}}

\def\bgrk#1{\mbox{{\boldmath $#1$ \unboldmath}}\!\!}
\def\eq#1{(\ref{#1})}
\def\H1{\widehat{H}_1}
\def\M{\hbox{\large \tt M}}
\def\dmat{\hbox{\large \tt d}}

\renewcommand{\i}{\ensuremath{\mathrm{i}}}
\newcommand{\e}{\ensuremath{\mathrm{e}}}
\renewcommand{\d}{\ensuremath{\mathrm{d}}}
\newcommand{\pd}{\partial}

\title{Probing quantum and thermal noise in an interacting many-body system}

\author{S.~Hofferberth}
 \affiliation{Atominstitut der \"Osterreichischen Universit\"aten, TU-Wien, Stadionallee 2, 1020 Vienna, Austria}
 \affiliation{Physikalisches Institut, Universit\"at Heidelberg, Philosophenweg 12, 69120 Heidelberg, Germany}
 \affiliation{Department of Physics, Harvard University, Cambridge, MA 02138, USA}
\author{I.~Lesanovsky}
 \affiliation{Physikalisches Institut, Universit\"at Heidelberg, Philosophenweg 12, 69120 Heidelberg, Germany}
 \affiliation{Institut f\"{u}r Theoretische Physik, Universit\"{a}t Innsbruck, Technikerstr. 21a, 6020 Innsbruck, Austria}
\author{T.~Schumm}
 \affiliation{Atominstitut der \"Osterreichischen Universit\"aten, TU-Wien, Stadionallee 2, 1020 Vienna, Austria}
\author{A.~Imambekov}
 \affiliation{Department of Physics, Harvard University, Cambridge, MA 02138, USA}
 \affiliation{Department of Physics, Yale University, New Haven, CT 06520,USA}
\author{V.~Gritsev}
\author{E.~Demler}
 \affiliation{Department of Physics, Harvard University, Cambridge, MA 02138, USA}
\author{J.~Schmiedmayer}
 \affiliation{Atominstitut der \"Osterreichischen Universit\"aten, TU-Wien, Stadionallee 2, 1020 Vienna, Austria}
 \affiliation{Physikalisches Institut, Universit\"at Heidelberg, Philosophenweg 12, 69120 Heidelberg, Germany}

\begin{abstract}
The probabilistic character of the measurement process is one of
the most puzzling and fascinating aspects of quantum mechanics. In
many-body systems quantum mechanical noise reveals non-local
correlations of the underlying many-body states. Here,
we provide a complete experimental analysis of the shot-to-shot
variations of interference fringe contrast for pairs of
independently created one-dimensional Bose condensates. Analyzing
different system sizes we observe the crossover from thermal to
quantum noise, reflected in a characteristic change in the
distribution functions from Poissonian to Gumbel-type, in
excellent agreement with theoretical predictions based on the
Luttinger liquid formalism. We present the first experimental
observation of quasi long-range order in one-dimensional atomic
condensates, which is a hallmark of quantum fluctuations in
one-dimensional systems. Furthermore, our experiments constitute
the first analysis of the full distribution of
quantum noise in an interacting many-body system.
\end{abstract}

\date{\today}
\maketitle

\section{Introduction}
The probabilistic nature of Schr\"odinger wave functions and the
uncertainty principle are crucial aspects of the modern
understanding of quantum matter. Starting with the famous
Bohr-Einstein debates, intrinsic quantum mechanical noise has been
the subject of numerous discussions and controversies
\cite{Zurek1984}. The analysis of quantum and thermal noise is not
only important in the study of fundamental problems in physics but
also crucial for understanding the ultimate limits of optical and
electric detectors, sensors, and sources.

The understanding of noise in the non-classical states of light
facilitated the development of the theory of photodetection and
led to the foundation of quantum optics \cite{Zoller2004}. In
solid state systems current fluctuations were used to probe the
nature of electrical transport in mesoscopic electron systems
\cite{Levitov2003,Sukhorukov2007} and to investigate quantum
correlations and entanglement in electron interferometers
\cite{Samuelsson2004, Neder2007}.

In atomic physics, noise correlation analysis \cite{Altman2004},
was employed to study quantum states in optical lattices
\cite{Foelling2005,Rom2006}, pair correlations in
fermi gases \cite{Greiner2005}, the counting statistics in an atom
laser \cite{Ottl2005}, and the Hanburry-Brown-Twiss effect for
both bosons and fermions \cite{Schellekens2005,Jeltes2007}.

Interference experiments provide a different powerful tool which
allowed, for example, the study of macroscopic phase coherence
\cite{Andrews1997}, critical fluctuations \cite{Donner2007},
thermal fluctuations in elongated condensates \cite{Richard2003},
and the Berezinskii-Kosterlitz-Thouless transition in a
two-dimensional quantum gas \cite{Hadzibabic2006}. Recently it has
been suggested that the full statistics of fluctuations in the
contrast of interference fringes can be used to probe high order
correlation functions and reveal non-trivial phases of
low-dimensional condensates \cite{Polkovnikov2006,Gritsev2006}.

In this paper we develop this idea further and provide a complete
experimental analysis of the \emph{shot-to-shot variations} of the
interference fringe contrast for pairs of independently created
one-dimensional Bose condensates. For long system sizes we find
that both the average contrast and its variations are dominated by
thermal fluctuations. For smaller systems we demonstrate that the
distribution functions of fringe contrast provide unambiguous
signatures of quantum fluctuations. Earlier experiments with
one-dimensional condensates in optical lattices observed
manifestations of strong atomic interactions in the three body
recombination rate \cite{Tolra2004}, the total energy
\cite{Kinoshita2004}, and the momentum distribution of atoms
\cite{Paredes2004}. However the power-law nature of the
correlation functions, which is the hallmark of one-dimensional
quasi condensates, has never been observed before. Our experiments
provide the first experimental demonstration of quasi long-range
order in one-dimensional condensates. This is the smoking gun
signature of quantum fluctuations in systems of weakly interacting
ultracold atoms. More importantly, our work constitutes the first
measurements of the full distribution functions of quantum noise
in an interacting many-body system.

\section{Experimental procedure}
\begin{figure}
\includegraphics[angle=0,width=\columnwidth]{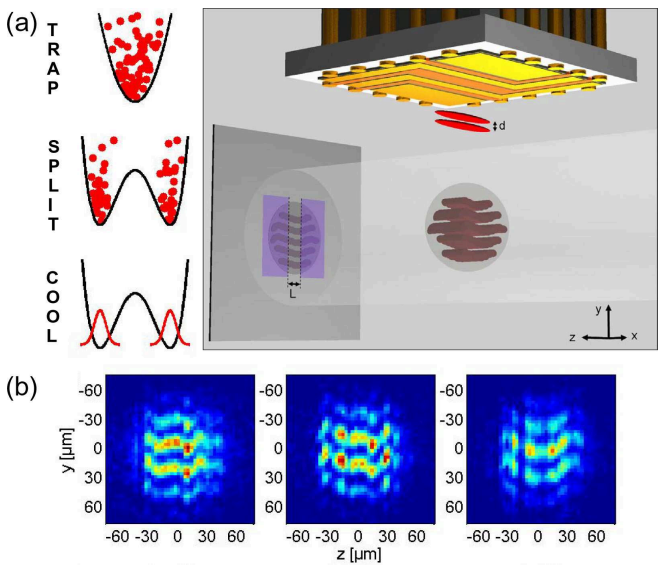}
\caption{\label{Fig:setup} Experimental setup and observed
interference patterns. \textbf{(a)} Two independent 1d Bose gases
are created by first splitting a single highly elongated magnetic
trap on an atom chip holding a thermal ensemble of atoms into a
double-well using radio-frequency-induced potentials. In a second
step the separate parts are evaporatively cooled to degeneracy,
producing two individual 1d condensates (schematic on the left).
The two systems are then simultaneously released from the trapping
potential and the resulting interference pattern is recorded with
standard absorption imaging. The vertical orientation of the
initial system is chosen so that the interference pattern can be
imaged along its transverse direction, parallel to the atom chip
(illustration on the right side). \textbf{(b)} Color coded images
of the resulting density patterns. The observed interference
fringes show a meandering over the length of the system
(z-direction), which is due to the local differences in relative
phase between the two original 1d condensates. Consequently, this
waviness of the patterns contains information about the phase
correlations in the individual condensates.}
\end{figure}

Our experiments are performed using two independent
one-dimensional quantum degenerate atomic Bose gases created in a
radio-frequency-induced micro-trap \cite{Schumm2005,
Hofferberth2006,Lesanovsky2006} on an atom chip
\cite{Folman2002,Fortagh2007}.  Each sample contains typically
$3000-5000$ atoms in the 1d regime \cite{Petrov2004,Bouchoule2007}, with both temperature $T$ and
chemical potential $\mu$ fulfilling $k_{B} T, \mu <h \nu_\perp$,
where $\nu_\perp = 3.0$\, kHz is the trapping frequency of the
harmonic transverse confinement. When the two independent 1d
condensates are released, they recombine in time-of-flight and the
resulting interference pattern is recorded using standard
absorption imaging (figure \ref{Fig:setup}a).

\begin{figure}
\includegraphics[angle=0,width=\columnwidth]{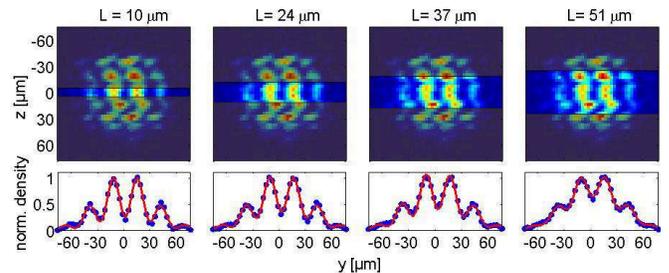}
\caption{\label{Fig:analysis}Analysis of the observed interference
patterns. For quantitative analysis, we integrate over central
slices of varying length $L$ of the density profiles in the
longitudinal direction  as indicated by the shading in the top
row, to obtain multiple transverse line density profiles. We then
extract the interference amplitude $|A_Q|$ by Fourier transforming
these profiles and extracting the Fourier coefficient
corresponding to the fringe spacing $Q$. To illustrate contrast
reduction with increasing $L$, the fringe patterns shown in the
bottom row are normalized. Modulated cosine fits to these profiles
then yield contrasts $C(L)$ which decrease with $L$. Note that the
interference amplitude $|A_Q|$ as defined in equation (\ref{C_Q})
is related to this contrast as $|A_Q| = n_{1d}\,L \times C$.
Consequently, as can be seen from equation
(\ref{avg_contrast_L_dependence}), $|A_Q(L)|$ increases with $L$,
whereas the contrast $C(L)$ decreases with $L$.}
\end{figure}

Examples of the observed fringe patterns are shown in figure
\ref{Fig:setup}b. While the interference patterns have high local
contrast, the interference fringes as a whole are not straight
lines. This meandering character of the interference patterns
shows that the relative phase between the two condensates is not
constant but fluctuates from point to point. These phase
variations originate from both quantum and thermal fluctuations in
the original 1d condensates and reflect the non-mean-field
character of low-dimensional systems. Integrating local
interference patterns over a finite length $L$ leads to summing
interference patterns which are not in phase with each other and
results in a reduction of the total fringe contrast (figure
\ref{Fig:analysis}). This reduction of the interference contrast,
and its statistical fluctuations, contains important information
about the phase correlations of the individual 1d condensates and
is the main quantity addressed in this work.

\section{Theoretical model}
Before we proceed to the statistical analysis of the experimental
data we present a quick overview of the theoretical foundations of
our study. The analysis is performed in two steps. In the first
step we analyze the \emph{average amplitude} of interference
fringes as a function of the integration length $L$, the second
step then analyzes the quantum and thermal noise contributions to
the \emph{shot-to-shot fluctuations in the average contrast}

Our interference patterns show a periodic density modulation at
the interference wave vector $Q=md/\hbar t$, where $m$ is the mass
of the atoms, $d$ is the in-trap separation of the two 1d systems,
and $t$ is the expansion time. Assuming ballistic expansion, the
complex amplitude of this density modulation after integration
over a length $L$ is given by \cite{Polkovnikov2006,Imambekov2007}
\begin{eqnarray}
A_Q(L)= \int_{-L/2}^{L/2} dz\, a^\dagger_1(z) a_2(z).
\label{C_Q}
\end{eqnarray}
Here $a_{1,2}$ are the boson annihilation operators within the two
original one-dimensional condensates before the expansion. The phase of
$A_Q(L)$ describes the position of interference fringes and is
determined by the relative phase between the two condensates
averaged between -L/2 and +L/2 .

Since we perform interference experiments with \emph{independent}
condensates, the phase of $A_Q$ is random, and the expectation
value $ \langle A_Q \rangle$ is zero. This does not imply the
absence of fringes but shows the unpredictable random phase in
individual interference patterns
\cite{Castin1997,Javanainen1997,Leggett2001}. Consequently, to
study the \emph{contrast statistics} of the interference patterns
one has to consider the quantity $\langle |A_Q(L)|^2 \rangle$.
This quantity is independent of the overall phase difference but
is strongly affected by phase twisting within each condensate.

In the case of ideal non-fluctuating condensates one expects to
find perfect contrast for any size of the system. This implies
$\langle |A_Q(L)|^2 \rangle \propto L^2$. In the opposite regime
of short range phase correlations with finite  correlation length
$\xi_\phi$, the net interference pattern comes from adding up
fringes in $L/\xi_\phi$ uncorrelated domains. In this case the net
interference pattern is strongly suppressed and appears only as a
square root fluctuation, $\langle |A_Q(L)|^2 \rangle \propto L
\xi_\phi$.

More precisely, $\langle |A_Q(L)|^2 \rangle$ is determined by the
integral of the two point correlation function:
\begin{eqnarray}
 \label{C2_Q}
 \langle | A_Q(L) |^2 \rangle = & & \int_{-L/2}^{L/2} dz_1 \int_{-L/2}^{L/2} dz_2  \\
 & & \langle a_1^\dagger(z_1) a_1(z_2) \rangle \langle a_2^\dagger(z_2) a_2(z_1) \rangle. \nonumber
\end{eqnarray}

A special feature of one-dimensional systems of interacting bosons
is the dramatic enhancement of fluctuations. Even at $T=0$, true long-range order is not possible
and only quasi-condensates with a power-law decay of the
correlation function $\langle a^\dagger(z_2) a(z_1) \rangle$
exist \cite{Popov1983}. At finite temperatures one finds exponential decay of the
correlation function for distances $|z_2-z_1|$ exceeding a thermal
correlation length $\xi_{\phi}(T)$ \cite{Petrov2004}.

To adequately describe these systems, a beyond mean-field theory is required. A powerful non-perturbative
approach which describes the long distance behavior of the
correlation functions of one-dimensional systems is the Luttinger
liquid theory (see methods and Refs.
\cite{Haldane1981,Giamarchi2003,Cazalilla2004}), on which we base
our further analysis.

Using a standard expression for the two point correlation function
in Luttinger liquid theory we obtain
\begin{eqnarray}
 \label{avg_contrast_L_dependence}
 \langle|A_Q(L)|^2\rangle=C_1 \xi_h \; L + C_2 \left(\frac{L}{\xi_h}\right)^{2-1/K} f(\frac{\xi_{\phi}(T)}{L},K).
\end{eqnarray}

Here, $K \approx \pi \hbar \sqrt{\frac{n_{1d}}{g m}}$ is the
Luttinger parameter for the weakly interacting 1d Bose gas, with
$n_\text{1d}$ being the 1d line density, $g = 2 h \nu_\perp a_s$
the effective 1d coupling constant, and $a_s$ being the s-wave
scattering length. $\xi_h = \frac{\hbar}{\sqrt{m g n_{1d}}}$ is
the healing length and $\xi_{\phi}(T) = \frac{\hbar^2 n_{1d} \pi
}{m k_B T}$ is the thermal correlation length of the 1d
condensates (for the weakly interacting regime), while $C_1$ and
$C_2$ are numerical constants of order unity. The function $f(x,
K)$ is given by
\begin{eqnarray}
 \label{f_integral}
 f(x, K)=\int_0^1\int_0^1 du dv \left(\frac{\pi}{x \sinh(\frac{\pi |u-v|}{x})}\right) ^{1/K}.
\end{eqnarray}
Note that the finite number of particles can in principle lead to
corrections to equation (\ref{f_integral})
\cite{Polkovnikov2007,Imambekov2007}. We checked that for our
parameters this shot noise is of no importance even for the
smallest $L$ we investigate.

The first term in equation (\ref{avg_contrast_L_dependence}) is
generally small. It describes contributions from the short
distance part of the correlation function, which is not sensitive
to thermal fluctuations (provided that $T < \mu$ as in our system)
or the non-trivial effects of quantum fluctuations in 1d systems.
The second term shows how quantum and thermal fluctuations in the
1d condensates affect $\langle |A_Q(L)|^2 \rangle$.

Let us first discuss the case of low temperatures and/or short
system sizes ($L/\xi_{\phi}(T) \leq 1$). Here we only need to
consider quantum fluctuations, which originate from interactions
between atoms. Non-interacting bosons at zero temperature have no
phase fluctuations, their interference pattern exhibit perfect
contrast, leading to $\langle |A_Q(L)|^2 \rangle \propto L^2$. In
the other extreme, impenetrable bosons (Tonks-Girardeau gas) have
very strong fluctuations and their interference pattern
corresponds to short range correlations, $\langle |A_Q(L)|^2
\rangle \propto L $ (see discussion above, eq. (2)). For finite
interaction strength we find something in between, resulting in
the scaling $\langle |A_Q(L)|^2 \rangle \propto L^{2-1/K}$.

Finite temperature introduces thermal fluctuations, which create
phase fluctuations with a temperature dependent correlation length
$\xi_{\phi}(T)$. When $L > \xi_{\phi}(T)$ thermal fluctuations
dominate and the interference amplitude scales as $\langle
|A_Q(L)|^2 \rangle \propto L \xi_{\phi}(T)$.

The experimentally observed interference patterns provide us with
more information than just the average value $\langle |A_Q(L)|^2
\rangle$. As a second step of our analysis we consider the
shot-to-shot fluctuations of individual measurements, which are
characterized by the higher moments $\langle |A_Q|^{2n} \rangle$
and ultimately by the entire distribution function
$W(|A_Q(L)|^2)$. For visualizing the shot-to-shot fluctuations of
the interference amplitude, it turns out to be convenient to
consider the normalized variable $\alpha(L) = |A_Q(L)|^{2}/\langle
|A_Q(L)|^{2} \rangle$ and its distribution function
$W(\alpha(L))$. The importance of the higher moments $\langle
|A_Q|^{2n} \rangle$ is that they are directly related to the
higher order correlation functions of the 1d interacting Bose gas
\cite{Polkovnikov2006}.

In the following, we only summarize the scaling of the
distribution function, a formal theoretical approach is discussed
in the methods. In the case of non-interacting ideal condensates,
i.e. perfect interference patterns in each measurement, the
distribution function $W(\alpha(L))$ approaches a delta function.
When interactions are weak but finite, we expect a narrow
distribution of width $1/K$ and $W(\alpha(L))$ to approach a
universal Gumbel distribution \cite{Gumbel1958}. In the limit of
long integration lengths, $L \gg \xi_{\phi}(T)$, thermal
fluctuations dominate. As discussed above, in this case the net
interference pattern comes from adding local interference patterns
from many uncorrelated domains resulting in the distribution
function being Poissonian. For integration lengths comparable to
$\xi_{\phi}(T)$ both quantum and thermal fluctuations are
important. In this regime we expect $W(\alpha(L))$ to show a
double peak structure, with the peak at small amplitudes coming
from the thermal noise and the peak at finite amplitude from
quantum noise.

\section{Analysis of the average interference amplitude}
\begin{figure}
\includegraphics[angle=0,width=\columnwidth]{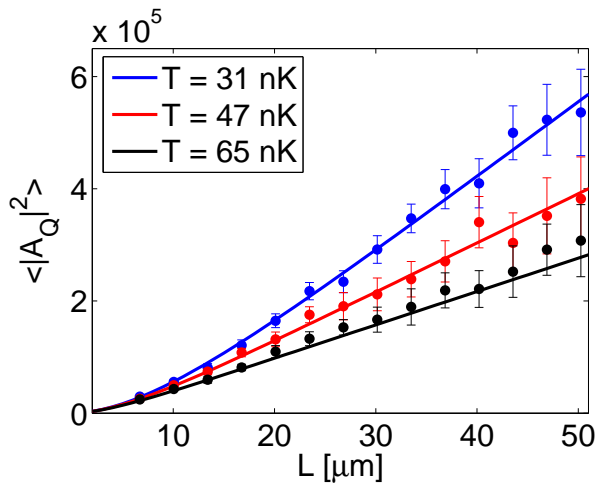}
\caption{\label{Fig:averages}Length dependence of the average
contrast. The data points show the measured $\langle|A_Q
|^2\rangle$ for three different temperatures $T$. Error bars
indicate s.e.m. Each data point contains 50 individual
interference measurements. The solid lines are fits of equation
(\ref{avg_contrast_L_dependence}) to the data with $T$ as free
parameter (see methods) and parameters $C_{1}=0.91$ and
$C_{2}=0.32$. $K$ and $\xi_h$ are determined independently from
measurements of $n_{1d}$ and $\nu_\perp$. }
\end{figure}

We now turn to the analysis of our experimental data, starting
with the average interference amplitude square $\langle |A_Q|^2
\rangle$ and its variation with system size $L$.

To extract $|A_Q|$ from the observed interference patterns, we
first obtain transverse density profiles integrated over the
longitudinal direction for different length $L$, as shown in
figure \ref{Fig:analysis}. We then fit a cosine function with a
Gaussian envelope to the resulting fringe profiles to extract the
relative phase and the interfering amplitude $|A_Q|$ as functions
of $L$ (see methods). To ensure homogeneous 1d density, we
restrict our analysis to the central $50\%$ of the system, where
$n_{1d}(z)$ varies at most by $\sim 10\%$ for the largest $L$
considered. This modulation is neglected and we obtain a single
value for $n_{1d}$ by averaging over the atomic density in this
center region.

Figure \ref{Fig:averages} shows the experimentally observed
average interference amplitude squares $\langle |A_Q|^2 \rangle$
for three different temperatures, with the density $n_{1d} = 50
\,\mu$m$^{-1}$ and the transverse trapping frequency $\nu_\perp =
3.0$ kHz (Luttinger parameter $K=42$ and healing length $\xi_h =
0.3 \,\mu$m) identical for all three data sets. The higher
temperature data sets are obtained by waiting for different times
after the initial preparation of the two condensates. During this
waiting time, the system heats due to residual noise in the
magnetic trapping fields.

To compare measurement and theory, we fit the function
(\ref{avg_contrast_L_dependence}) to the experimental data (figure
\ref{Fig:averages}) with the temperature $T$ as a free parameter
(see methods). We find the functional behavior of the measured
contrasts to be in very good agreement with the theoretical
predictions. This is of particular interest as the shape of these
curves is determined by both the quantum and thermal contributions
to the average contrast, as discussed above. For integration
length longer than 20-30\,$\mu$m we observe a linear dependence of
$\langle |A_Q(L)|^2 \rangle$ on $L$. This corresponds to the $L >>
\xi_{\phi}(T)$ regime where thermal fluctuations dominate.

For shorter segment lengths, quantum fluctuations are important.
However the analysis presented in figure \ref{Fig:averages} is not
sufficient to make the case for quantum fluctuations. The
Luttinger parameter for our system is $K=42$, and it is impossible
to observe the $L^{-1/K}$ correction to the ideal case
(noise-free) power law $L^2$ in the limited range of lengths
available. From figure \ref{Fig:averages} we can not prove that
fluctuations are present at all for such short system sizes. We
will address this in the next part of our analysis by
demonstrating that quantum fluctuations manifest unambiguously  in
the \emph{shot-to-shot fluctuations} of $ |A_Q(L)|^2 $, rather
than in the $\langle |A_Q(L)|^2 \rangle$ average value.

>From the fits we obtain the temperatures $T\,=\,31,\, 47,\,
65$\,nK for $0,\, 50,\, 100$\,ms waiting time, respectively. As
expected, the greater the waiting time, the higher the temperature
of the system. We note here that this method measures the
temperature of collective excitation in the condensate. We cannot
confirm that this temperature is identical to that of the residual
thermal atoms in the trap. Reliable detection of the thermal
background is possible only down to $T \approx 80$\,nK in our
setup.

The contrast method we present here can be used to measure the
temperature of collective modes of 1d Bose gases at extremely low
temperatures and small atom numbers, suggesting the usefulness of
this method for precise thermometry of 1d condensates when
conventional methods fail.

\section{Analysis of the full distribution functions of interference amplitude}
\begin{figure}
\includegraphics[angle=0,width=\columnwidth]{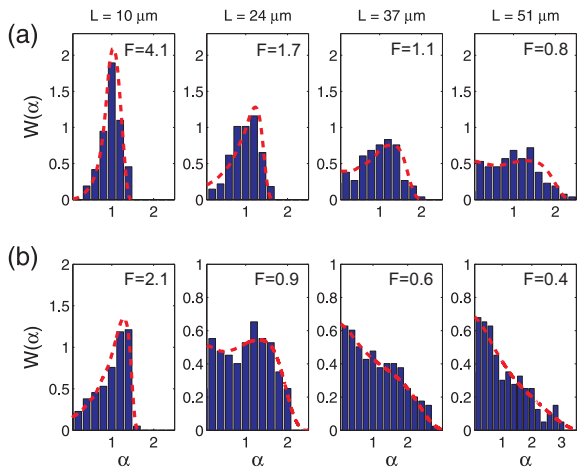}
\caption{\label{Fig:distributions}Distribution functions of the
measured interference contrasts for different lengths $L$.
\textbf{(a)} The length-dependent normalized interference
contrasts $\alpha(L) = |A_Q(L)|^{2}/\langle |A_Q(L)|^{2} \rangle$
of 170 individual experimental realizations with identical
parameters ($n_{1d}=60 \,\mu$m$^{-1}$, $\nu_\perp=3.0$\,kHz,
$K=46$) are displayed as histograms. The red curves show the
corresponding calculated distributions for $T=30$\,nK
($\xi_{\phi}(T)=4.1 \,\mu$m). \textbf{(b)} Histograms of 200
individual measurements with the same parameters as in
\textbf{(a)}, but higher temperature $T=60$\,nK
($\xi_{\phi}(T)=2.1 \,\mu$m). For both sets we observe very good
agreement between experiment and theory. In particular, the
predicted change of overall shape of the distribution functions
from single peak to Poissonian with decreasing $F=\xi_{\phi}(T)/L$
(increasing $L$ and $T$) is very well reproduced by the
experimental data.}
\end{figure}

We now analyze the full information contained in the
statistics of the interference contrast.

In figure \ref{Fig:distributions}, we show histograms of the
measured distributions $W(\alpha(L))$ for four different length
scales $L$ and two different temperatures, $T=30$\,nK (upper row,
figure \ref{Fig:distributions}(a)) and $T=60$\,nK (lower row,
figure \ref{Fig:distributions}(b)), obtained using the contrast
average method discussed in the previous section. Both data sets
were obtained from 1d condensates with $n_{1d}=60 \,\mu$m$^{-1}$,
$\nu_\perp=3.0$\,kHz, and $K=46$.

For a detailed comparison between measurement and theory, we
numerically calculate the distribution functions for the
corresponding experimental parameters for each histogram (see
methods and Ref. \cite{Imambekov2006}). We emphasize that once we
know the temperature of the system, there are no free parameters
remaining in this analysis. It is truly remarkable that the
experimentally measured distribution functions are in such
excellent agreement with the theoretical prediction (figure
\ref{Fig:distributions}). In particular we clearly observe the
transition from the regime dominated by quantum fluctuations for
small L at low temperature to the one dominated by thermal
fluctuations for large L and higher temperature.

More quantitatively, the shape of the distribution functions is
determined by a single dimensionless parameter $F=K\xi_{T}/L$. For
large $F$, accessed either at low temperature $T$ or small system
length $L$ we find a single \emph{asymmetric} peak in the
distribution function $W(\alpha)$, resulting from a universal
extreme-value statistics Gumbel distribution \cite{Gumbel1958}
which is a smoking gun signature of quantum fluctuations and the
power-law behavior of the correlation functions.

The Gumbel distribution has a characteristic asymmetric shape and
typically appears when describing rare events such as stock market
crashes or earthquakes, which go predominantly in one direction.
This suggests that suppression of the contrast of interference
fringes due to quantum fluctuations is dominated by rare but
strong fluctuations of the phase of the bosonic fields
$a_{1,2}(z)$. Most of the time we find very small phase meandering
which does not affect the contrast significantly. Only
occasionally there is a strong fluctuation leading to a noticeable
decrease of the interference fringe contrast.

For longer system sizes $L$ and higher temperatures $T$, we
observe that the distribution functions become Poissonian,
characteristic for the dominance of thermal fluctuations and the
exponentially decaying correlations. The system can be thought of
as consisting of domains of size $\xi_{\phi}(T)$, with
uncorrelated phases in each of the domains. In this case, adding
up interference amplitudes (complex numbers) is similar to
performing a random walk in two dimensions and the total amplitude
(distance travelled) is proportional to the number of steps,
resulting in the net interference contrast being proportional to
$1/\sqrt{L}$.

Finally, for intermediate lengths and temperatures, we observe the
formation of a double-peak structure in the distribution
functions, characterized by a peak at zero, originating from
thermal noise and a peak at finite amplitude $\alpha$ originating
from quantum noise \cite{Imambekov2006,Imambekov2007}. In this
crossover regime the relative effects of quantum fluctuations are
diminished but not completely suppressed, so that both quantum and
thermal fluctuations are of importance in determining the shape of
the distribution functions. We find very good agreement between
experiment and theory also in this crossover regime.

\section{Discussion and Summary}

It is interesting to note that the above analysis is based on the
Luttinger liquid theory of interacting bosons in one dimension
which is accurate for calculating the long-range part of the
correlation functions but does not capture the short distance part
on the scale of the healing length $\xi_h$.  For system size $L
\gg \xi_h$ it is this long distance part of the correlation
functions that gives the dominant contribution to the integrals
determining the interference amplitude (see equation
(\ref{C2_Q})). This sensitivity to the long distance part of the
correlation functions is the unique feature of interference
experiments, which makes them a powerful tool for analyzing
quantum and thermal fluctuations in low-dimensional condensates.

Alternative approaches, such as measurements of density
fluctuations in expanding condensates \cite{Dettmer2001} are probing correlation
functions on the scale of the healing length. This short range
part of the correlation function is hardly sensitive to the quasi
long-range nature of quantum fluctuations in one-dimensional
systems. This makes it difficult to observe quantum effects by direct measurement of density
fluctuations: they reveal the role of interactions
\cite{Esteve2006} but the transformation of short range
correlations into density fluctuations masks the quantum
correlations.



In summary, we have studied quantum and thermal noise in 1d
systems of interacting quantum degenerate bosons using the full
distribution function of the interference amplitude. The
shot-to-shot fluctuations in the contrast contain information,
which can be related to high-order correlation functions of
the 1d system. Our results provide the first experimental
measurements of the full distribution function of quantum
noise in an interacting many-body system. By analyzing these
distribution functions we provide the first direct experimental
evidence of quasi long-range order in one-dimensional condensates.
The remarkable agreement between our experimental findings and
theoretical predictions based on the Luttinger liquid model
provides a confirmation of this theoretical approach as an
effective low-energy theory of interacting bosons in one
dimension. This demonstrates the power of quantum noise analysis
in studying strongly correlated many-body systems.

We expect our experiments to pave the way for other methods of
characterizing many-body systems using the analysis of quantum
noise like particle number fluctuations \cite{Bruder2007} and spin
noise \cite{Lewenstein2007,Cherng2007,Eckert2007}. From the point
of view of analyzing systems with strong interactions and
correlations, this should allow cold-atom experiments to provide a
complementary and different perspective to that provided by
electron systems.

\section{Methods}

\subsection{Preparing two independent 1d condensates on an atom chip}
We start the experiment with a thermal ensemble of  $\sim 10^5$
Rb$^{87}$ atoms in the $|F=2,m_{F}=2>$ state at a temperature $T
\approx 5 \,\mu$K in a single highly elongated magnetic trap on an
atom chip \cite{Folman2002,Fortagh2007}.  This initial sample is
prepared using our standard procedure of laser cooling, magnetic
trapping, and evaporative cooling \cite{Wildermuth2004}. The
initial trapping configuration is then deformed along the
transverse direction into a highly anisotropic double-well
potential by means of radio-frequency-induced adiabatic potentials
\cite{Schumm2005, Lesanovsky2006}. In particular, we employ the
rf-trap setup presented in ref. \cite{Hofferberth2006} where the
combination of two radio-frequency (rf) fields generated by wires
on the atom chip allows the realization of a compensated symmetric
double-well potential in the vertical plane. The final cooling of
the two separated ensembles leading to the two 1d condensates is
achieved by performing forced evaporative cooling in the dressed
state potential \cite{Hofferberth2007}. We observe the onset of
quantum degeneracy at $T\approx 400$\,nK in each of the two
potential tubes.

The potential barrier between the two systems is controlled by the
amplitude of the rf fields and the gradient of the static magnetic
trap \cite{Lesanovsky2006}. We realize a barrier height $V \approx
k_B \times 4 \,\mu$K to ensure a complete decoupling of the two
systems during the final cooling stage.

After cooling and a relaxation time of $300$\,ms to ensure each
system is in equilibrium (a constant rf knife is kept on during
this time to prevent heating), each potential tube contains
$3000-5000$ atoms at typical temperatures $T < 100$\, nK. The
atoms are trapped in a strong transverse harmonic confinement of
$\nu_\perp \sim 3.0$\,kHz at a distance of $75 \,\mu$m from the
atom chip surface. Each individual degenerate atomic ensemble is
in the one-dimensional regime, with both temperature $T$ and
chemical potential $\mu$ fulfilling $k_{B} T, \mu <h \nu_\perp$
\cite{Petrov2004,Bouchoule2007}.

\subsection{Measuring the interference pattern}
We observe the interference pattern created by the two expanding,
overlapping atomic clouds using standard absorption imaging. For
the vertical double-well orientation used in the experiments, the
observed interference fringes in the atomic density are
horizontal, parallel to the atom chip surface. This enables us to
image the interference pattern along the transverse direction of
the system. The employed imaging system has a spatial resolution
of $3.4 \,\mu m$ and a noise floor of $\sim 3$ atoms per 3x3\,$\mu
m$ pixel.

\subsection{Extracting the fringe amplitude from interference patterns}
From a single interference image we obtain line profiles for
different $L$ by integrating the two-dimensional absorption image
over various lengths along the longitudinal direction of the
system. The obtained line densities are then Fourier transformed,
and we extract $A_Q$ as the value of this Fourier transform at the
wave vector $Q$ corresponding to the observed fringe spacing. This
spacing is determined from fitting the interference patterns with
a cosine function with a Gaussian envelope plus an unmodulated
Gaussian to account for the contrast reduction. The free
parameters of these fits are the relative phase $\theta$, the
contrast, and the fringe spacing. The width, amplitude, and center
position of the total cloud are determined independently from a
Gaussian fit to the full integrated density pattern of the central
area of each image. Note that the absolute value of the
interference amplitude $|A_Q|$ (as defined in eq. \ref{C_Q}) and
the contrast $C$ are related as $|A_Q|= n_{1d}\,L \times C$.

\subsection{Average interference amplitude fits}
To compare the experimentally observed length dependence of the
average interference amplitude to theory, we perform a least
square fit of the theoretical prediction (equation
(\ref{avg_contrast_L_dependence})) to the data. Since the line
density $n_{1d}$ is extracted directly from the absorption images
and the transverse trapping frequency $\nu_\perp$ is measured
precisely by parametric heating experiments, the only unknown
experimental parameter is the temperature $T$. We include a global
additive fit parameter to account for contrast reduction due to
technical aspects such as finite imaging resolution and focal
depth.

\subsection{Luttinger liquid}
A one-dimensional gas of ultra-cold bosonic atoms can be described
by the  Lieb-Liniger model of bosons interacting via a point-like
repulsion \cite{Olshanii1998, Lieb1963, Kinoshita2004}. The
effective approach to the Lieb-Liniger model, capturing the
long-distance behavior of all correlation functions is known as
the Luttinger liquid formalism
\cite{Haldane1981,Giamarchi2003,Cazalilla2004}. The essence of
this approach is to represent the original bosonic field in terms
of the two phase fields
$a(z)=(n_{1d}+\partial\theta(z))^{\frac{1}{2}}e^{i\phi(z)}$ and
keep only the terms quadratic in $\phi(z),\theta(z)$ in the
Hamiltonian. The resulting theory has linear spectrum of bosonic
sound waves and shows algebraic decay of all correlation functions
at zero temperature (e.g $<a^{\dag} a>\sim |z_{1}-z_{2}|^{1/2K}$)
and exponential decay for finite temperature (e.g.
$<a^{\dag}(z_{1})a(z_{2})>\sim n_{1d}
[\pi/(\xi_{T}n_{1d}\sinh(\pi(z_{1}-z_{2})/\xi_{T})]^{1/2K}$.
Here $K$ is the fundamental parameter of the theory, the so-called
Luttinger parameter and $\xi_T=\xi_{\phi}(T)/K$. The last expression
applies down to the short distance cut-off given by the healing
length. The value of $K$ is uniquely determined by the dimensionless
ratio characterizing the original microscopic model:
$\gamma=mg/\hbar^{2} n_{1d}$, where $g$ is one-dimensional
interaction strength. In the weakly interacting regime studied here,
$K\approx\pi/\sqrt{\gamma}$. Recent analysis showed
\cite{Caux2006,Caux2007} that the Luttinger liquid formalism
provides extremely accurate description of the correlation functions
of the Lieb-Linger model for both long distances and distances just
beyond the healing length.

\subsection{Calculation of the distribution functions}
Computation of  the distribution functions requires, in principle,
the knowledge of all moments of the interference fringes
amplitude. One approach to overcome this problem of moments was
introduced in Ref. \cite{Gritsev2006}, where methods of conformal
field theory and special properties of exactly-solvable models
were used to compute the distribution function for periodic
boundary conditions at zero temperature. Another method, which
allows to compute the distribution functions for all boundary
conditions, arbitrary temperature, and in all dimensions
\cite{Imambekov2006} is based on the  mapping of the problem to a
generalized Coulomb gas model and a related problem of fluctuating
random surfaces (for a review, see Ref. \cite{Imambekov2007}).
This is the approach which we use in our analysis.

The  full distribution function $W(\alpha)$ is  defined by the
normalized moments of the interference fringe contrast as
\begin{equation}
\langle \alpha^m \rangle=\langle | A_Q |^{2m} \rangle / \langle |
A_Q |^{2} \rangle^m = \int_0^{\infty} W(\alpha)\alpha^{m}d\alpha.
\end{equation}
Using Luttinger liquid theory, these moments can be expressed
\cite{Polkovnikov2006} as the micro-canonical partition functions
of the Coulomb gas of $2m$ particles
\begin{eqnarray}
 &\langle \alpha^m \rangle &= \int_0^1 ...\int_0^1 du_1...dv_m \\
 &\times&\exp\left[\textstyle{\frac{1}{K}}\underset{i<j}{\sum}\{G(u_i,u_j)
 + G(v_i ,v_j)\}- \frac{1}{K}\underset{i,j}{\sum} G(u_i,v_j)\right].\nonumber
 \label{Z2n}
\end{eqnarray}
Here $G(x,y)$ is the interaction potential, which precise form
depends on the geometry of the problem and the temperature. At
zero temperature, $G(x,y)=\log{|x-y|},$ while at nonzero
temperature $ G_T(x,y)=\log{\left(\frac{\xi_T}{\pi
L}\sinh{\frac{\pi|x-y|L}{\xi_T}}\right)}. $ Real and symmetric
$G(x,y)$ can be decomposed as $ G(x,y)=\sum_{n=1}^{n=\infty}G_n
\Psi_{n}(x) \Psi_{n}(y). $ Such decomposition  is similar to
diagonalization of a symmetric matrix by finding its eigenvectors
and eigenvalues. Eigenfunctions $\Psi_{n}(x)$ and eigenvalues
$G_{n}$ of the interaction potential $G(x,y)$ can be used to
construct the height variable
$h(x,\{t_{n}\})=\sum_{n}t_{n}T_{n}(x)-T_{n}(x)^2/2,$ where
$T_{n}(x)=\Psi_{n}(x)\sqrt{G_{n}/K},$ and $t_{n}$ are fluctuating
noise variables. Introducing $g(\{ t_{n}\})=\int
dx\exp[h(x,\{t_{n})\}],$ the distribution function can be written
as \cite{Imambekov2006,Imambekov2007} \beq\label{w}
W(\alpha)=\prod_{n=1}^{\infty}\frac{\int^{\infty}_{-\infty} dt_n
e^{-t_{n}^{2}/2}}{\sqrt{2\pi}}\delta[\alpha-g(\{ t_{n}\})g(\{
-t_{n}\})]. \eeq We compute this function using a Monte-Carlo
algorithm. Random variables $\{t_n\}$ are chosen from the Gaussian
ensemble, and one-dimensional integrals $g(\{ t_{n}\}), g(\{-
t_{n}\})$ are evaluated for each realization of $\{t_n\}.$
According to equation (\ref{w}), distribution function $W(\alpha)$
coincides with the distribution function of the product $g(\{
t_{n}\})g(\{-t_{n}\}).$ In the limit of large parameter $F$ the
equation (\ref{w}) can be evaluated explicitly to show that the
distribution approaches one of the extreme-value statistics
distributions (similar to a Gumbel form)
\cite{Imambekov2007,Gumbel1958}.

We acknowledge financial support from the European Union, through
the contracts MRTN-CT-2003-505032 (Atom Chips), Integrated Project
FET/QIPC `SCALA', FWF, NSF, Harvard-MIT CUA, AFOSR, Swiss NSF and
MURI. We thank S. Groth for fabricating the atom chip used in the
experiments and D. A. Smith for critical reading of the
manuscript.

\end{document}